# Acoustic Supercoupling in a Zero-Compressibility Waveguide

H. Esfahlani,[1,2,3,†] M. S. Byrne,[2,4,†] M. McDermott,[2] A. Alù[1,2,5,6,*]

[1]Photonics Initiative, Advanced Science Research Center, City University of New York, New York City, NY 10031, USA.

[2]Department of Electrical and Computer Engineering, The University of Texas at Austin, Austin, TX 78712, USA.

[3]Electrical Engineering Institute, École Polytechnique Fedérale de Lausanne (EPFL), CH-1015 Lausanne, Switzerland.

[4]Naval Surface Warfare Center Carderock Division, West Bethesda, MD 20817, USA.

[5]Physics Program, Graduate Center, City University of New York, New York City, NY 10026, USA.

[6]Department of Electrical Engineering, City College of New York, New York City, NY 10031, USA.

†Authors contributed equally to this work.
*To whom correspondence should be addressed (email: aalu@gc.cuny.edu)

**Abstract**

Funneling acoustic waves through largely mismatched channels is of fundamental importance to tailor and transmit sound for a variety of applications. In electromagnetics, zero-permittivity metamaterials have been used to enhance the coupling of energy in and out of ultranarrow channels, based on a phenomenon known as *supercoupling*. These metamaterial channels can support total transmission and complete phase uniformity, independent of the channel length, despite being geometrically mismatched to their input and output ports. In the field of acoustics, this phenomenon is challenging to achieve, since it requires zero-density metamaterials, typically realized with waveguides periodically loaded with membranes or resonators. Compared to electromagnetics, the additional challenge is due to the fact that conventional acoustic waveguides do not support a cut-off for the dominant mode of propagation, and therefore zero-index can be achieved only based on a collective resonance of the loading elements. Here we propose and experimentally realize acoustic supercoupling in a dual regime, using a compressibility-near-zero acoustic channel. Rather than engineering the channel with subwavelength inclusions, we operate at the cut-off of a higher-order acoustic mode, demonstrating the realization and efficient excitation of a zero-compressibility waveguide with effective soft boundaries. We experimentally verify strong transmission through a largely mismatched channel and uniform phase distribution, independent of the channel length. Our results open interesting pathways towards the realization of extreme acoustic parameters, and their implementation in relevant applications, such as ultrasound imaging, sonar technology, and sound transmission.

# Introduction

Over the past decade, significant attention has been paid to zero-index metamaterials, due their extreme opportunities for wave manipulation [1]. These materials can be described by governing equations that are temporally and spatially decoupled, due to the unusual physics enabled by near-zero constitutive parameters. In turn, these effects lead to peculiar scattering and propagation phenomena [2]-[3]. The vast majority of this research has been focused in the electromagnetic domain, including media with near-zero dielectric permittivity (epsilon-near-zero or ENZ) [2]-[16], near-zero magnetic permeability (mu-near-zero or MNZ) [17] or double-near-zero materials (epsilon-and-mu-near-zero or EMNZ) [18]-[19]. Recent attention has also been given to analogous phenomena in the field of acoustics [20]-[24].

Remarkable properties of zero-index metamaterials have been proposed and experimentally validated. For instance, these metamaterials have been used for cloaking [25]-[28], improving the directivity in radiation and scattering [3],[29]-[30], and in enhancing the transmission through geometrically mismatched channels, a phenomenon dubbed as *supercoupling* [4],[27],[31]-[32]. It was shown that supercoupling enables electromagnetic energy to be tunneled through narrow channels filled with permittivity-near-zero materials, regardless of their geometrical shape and bending [5]-[9]. The dual phenomenon, tunneling through a geometrically mismatched large channel filled with permeability-near-zero materials was also theoretically proposed [17] and experimentally verified for radio-frequency waves [27]. The phenomenon can be explained through transmission-line theory [10] as the compensation of the large geometric mismatch between different waveguide sections with the extreme impedance values in zero-index metamaterials. ENZ supercoupling has been also proposed at higher frequencies as an effective way to boost optical nonlinearities in plasmonic channels [14] and local density of states and quantum super-radiance [33].

These advances have motivated the recent interest in exploring the physics of acoustic metamaterials with near-zero material properties. For example, a space-coiled structure was used to assess a density-near-zero material for acoustic tunneling [20]. Using a waveguide populated with transverse membranes and side holes, which exhibits density-and-compressibility-near-zero properties, an acoustic leaky-wave antenna with broadside radiation was realized [21]-[22]. Additionally, a membrane-based acoustic metamaterial with near-zero-density was proposed as an angular filter that only transmits waves with near-zero incident angle [23].

The analogue of zero-permittivity in acoustics, for the realization of acoustic supercoupling, is density-near-zero metamaterials [24]. One approach theoretically showed that energy could be squeezed through ultranarrow acoustic channels by employing a waveguide filled with arrays of transverse membranes [24], which indeed realized an effective zero-density ultranarrow channel. However, challenges with visco-thermal loss and the accurate tuning of multiple membrane resonances have prevented the practical realization of density-near-zero acoustic supercoupling devices thus far. A waveguide loaded with Helmholtz resonators in the form of low-pass filters was shown to support compressibility-near-zero properties and uniform phase through an intermediate channel [34], but again its implementation in a supercoupling experiment would require extreme precision in the realization of these arrays of resonators.

A big advantage that has enabled the realization of supercoupling in electromagnetics has been the fact that conducting waveguides naturally support effective zero-index properties at the cut-off of their dominant mode of propagation. This phenomenon has enabled several demonstrations of electromagnetic supercoupling in electromagnetics without having to realize a metamaterial through periodic arrays of small inclusions, but simply operating a hollow waveguide at cut-off [7]-[10]. Unfortunately, conventional acoustic waveguides typically do not provide a cut-off for their dominant propagating mode, as these modes are longitudinal in nature. In the present work, however, we show that it is possible realize acoustic supercoupling in a hollow waveguide by exciting a higher-order mode at cut-off, providing an experimentally viable, simple geometry demonstrating effective zero compressibility and supercoupling for sound. Despite the use of simple materials with physically hard boundaries, we show that the excitation of a higher-order mode may synthesize effective soft-boundary waveguide channels that support a cut-off at finite frequency, and therefore enable this unusual tunneling phenomenon. This approach establishes new pathways for extreme acoustic metamaterials, cloaking, acoustic sensing, and wave patterning.

**Results**

In order to demonstrate the effect of supercoupling for sound in a simple waveguide geometry, we consider the configuration of Fig. 1, in which we sandwich an intermediate channel with large cross-sectional area $S_2$, length $L$ and modal impedance $Z_2$ between two narrow input/output channels, each with acoustic impedance $Z_1$ and much narrower cross-sectional area $S_1$. As derived in the Methods section, the reflection coefficient at the input port reads

$$\Gamma = \frac{(Z_2^2 - Z_1^2)\tan(k_z L)}{(Z_1^2 + Z_2^2)\tan(k_z L) + 2jZ_1 Z_2}, \qquad (1)$$

where $k_z$ is the wave number in the middle channel. Reflection is minimized when $\tan(k_z L) = 0$ or when $Z_2 = \pm Z_1$. The first condition corresponds to conventional Fabry-Perot resonances, which depend upon the length and geometry of the connecting channel. However, tunneling independent of the channel length, as expected in supercoupling phenomena, can be achieved at the impedance matching condition $Z_2 = Z_1$.

A straightforward way to realize this matching condition in waveguides with large geometrical mismatch, as in Fig. 1, is to consider a rectangular channel bounded by two hard boundaries and two soft boundaries. This waveguide does not support a mode at low frequencies, and the dominant propagating mode, as derived in the Methods section, has impedance

$$Z_2 = \frac{\omega \rho_0}{S_2 \sqrt{\left(\frac{\omega}{c_0}\right)^2 - \left(\frac{\pi}{a}\right)^2}}, \qquad (2)$$

where $\omega$ is the driving frequency, $\rho_0$ is the density of the filling medium, $c_0$ is the corresponding sound velocity, and $a$ is the distance between the two soft boundaries. In particular, the impedance becomes very large at the cut-off frequency of the dominant mode $f = \frac{c_0}{2a}$, when the term in the square root goes to zero. For values close to cut-off, impedance matching can be achieved, as the small value of the square root is compensated by $S_2 \gg S_1$, yielding zero reflection and full transmission independent of the channel length. In order to confirm this intuition, Fig. 2a shows full-wave simulations evaluating the transmission through this geometry varying the channel length $L$. For each length, the first peak in transmission corresponds to the supercoupling frequency, which arises near the cut-off frequency of the intermediate channel. The higher-frequency transmission peaks are due to Fabry-Perot resonances, which are largely dependent on the channel length. By increasing the length of the channel, the number of Fabry-Perot resonances increase for a fixed frequency spectrum, however the tunneling frequency remains nearly unchanged. Fig. 2b presents the corresponding phase of the pressure field along the channel at the supercoupling frequencies, and the inset shows the 2D field distribution for the different lengths considered. We observe completely uniform phase along the channel, independent of the length of the coupling channel, consistent with propagation in a zero-index material, and with the previous observations of supercoupling in ENZ media for electromagnetic waves [5]-[10].

Indeed, for the intermediate channel the effective constitutive parameters can be retrieved as discussed in the Methods section, yielding

$$\begin{cases} \rho_{eff2} = \rho_0 \\ \kappa_{eff2} = \dfrac{\kappa_0}{1 - \left(\dfrac{\pi c_0}{a\omega}\right)^2} \end{cases}, \quad (3)$$

where $\kappa_0$ is the bulk modulus of the filling material. These equations confirm that at the cut-off frequency $f = \dfrac{c_0}{2a}$, $\kappa_{eff}$ has a pole, the effective compressibility goes to zero, and the phase velocity $\dfrac{\omega}{k_z} = \sqrt{\dfrac{\kappa_{eff}}{\rho_{eff}}}$ becomes infinite. We conclude that the geometry in Fig. 1 with mixed hard and soft boundary walls, realizes the dual of zero-density supercoupling, enabling the compensation of large geometrical mismatch in the middle channel through compressibility close to zero. The electromagnetic analogue has been explored in [17],[27] as a wide channel filled by MNZ materials. Consistent with these works and similar phenomena for zero-density supercoupling [24], Fig. 2c shows that supercoupling tunneling and infinite phase velocity is preserved also when the connecting channel is bent in different configurations, an effect of the quasi-static nature of wave propagation associated with its infinite phase velocity.

So far, we have shown that it is possible to achieve the equivalent of zero-compressibility propagation and supercoupling in a waveguide with mixed hard and soft boundary walls, operating near its cut-off frequency. However, this configuration is hardly realizable in a realistic geometry. Interestingly, in the following we show that it is possible to achieve an analogous functionality exciting a waveguide with all hard boundaries, as in the case of a conventional acoustic waveguide filled by air, at the cut-off frequency of one of its higher-order modes. In the Methods section we indeed show that the effective constitutive parameters of the $(m,n) = (2,0)$ mode supported by a hard-wall waveguide indeed satisfies a similar expression as in (3), ensuring a zero-compressibility condition near its cut-off.

The difference compared to the soft-hard waveguide in Fig. 2 is the presence of other modes in the waveguide, including the dominant mode $(m,n) = (0,0)$, which has no cut-off. However, we notice that these other modes, not being operated near the cut-off frequency, are badly mismatched to the input and output waveguides, due to the large geometrical mismatch. Therefore, their coupling to the input signal is negligible. In other words, the transition between different waveguides can be treated as a multi-port network, and the impinging energy naturally couples to the higher-order mode at cut-off, given that the impedance is conserved. In our geometry, we drive the middle channel at its center, totally preventing the excitation of odd-order modes because of symmetry. The dominant (0,0) mode is not excited due to the large impedance mismatch, and therefore all the energy can flow unperturbed into the second-order mode at the CNZ frequency.

Fig. 3 verifies this prediction in full-wave simulations of a hard-walled acoustic waveguide. Quite surprisingly, we retrieve very similar functionality in this configuration, relying on a simple hollow waveguide driven at the cut-off frequency of its (2,0) mode, $f = c_0 / a$. We can see that the spatial phase distribution of Fig. 3b matches well the one of Fig. 2b in the center of the channel, with infinite phase velocity throughout the channel independent of its length. In the transverse plane of the waveguide we observe a phase flip by $\pi$, associated with the modal distribution of the excited higher-order mode. Interestingly, the location of the flip is exactly the effective location of a soft-wall boundary, consistent with the results in Fig. 2. This concept may be useful to design effective soft-wall waveguide structures using materials that are not substantially softer than the filling medium. Also in this scenario, due to the quasi-static nature of the pressure field near the CNZ frequency, the acoustic wave tunneling occurs regardless of the shape of the channel, as demonstrated in Fig. 3c. In this case, the effective soft boundary also follows adiabatically the bending profile.

Fig. 4 presents our experimental verification of CNZ supercoupling in a channel with variable length operated near the cut-off frequency of its (2,0) mode. We built the intermediate channel using an off-the-shelf steel welder's tool box, in which a wooden wall was shifted in different positions to change the length of the channel in real time. The measured data capture well the physics of the problem: impedance matching and near-zero phase delay in transmission, independent of the length of the intermediate channel, are verified experimentally, despite the fact that the walls of our waveguide are not ideally hard (details of the experimental setup are provided in the Methods section). To aid in comparing numerical results to experiment, the model in Fig. 4 treated the boundaries of the coupling channel as elastic shells, which are capable of radiating sound into the surrounding air. This model also allowed for coupling of acoustic energy into vibrational modes of the walls, and for dissipation due to loss within the material (with $\tan \delta = 0.01$ for steel). This more realistic condition resulted in predicted transmission amplitudes with 8-13 dB of loss (depending upon length), whereas the experimental results showed slightly larger loss levels. This agreement in transmission is good, considering the uncertainties in material properties and variation of the intermediate channel geometry from an ideal rectangular prism. The data displayed lower-Q resonances than in the ideal scenario with hard walls, as expected, but confirmed nearly infinite phase velocity, and a tunneling frequency nearly independent of the channel length.

**Discussion**

In this work, we have presented theoretical and experimental validation of a straightforward way of realizing zero compressibility acoustic wave propagation in waveguides, by exciting a higher-order mode at its cut-off frequency. We used this unusual propagation regime to realize the supercoupling phenomenon for sound, enabling tunneling of energy through largely mismatched waveguide geometries. Our theoretical results capture well the physics behind this anomalous tunneling, and our experiments confirm large phase velocity and anomalous transmission independent of the channel length. The small discrepancies between measurements and numerical predictions can be explained by irregular geometry and uncertainties in the involved material properties of the off-the-shelf toolbox used for the middle channel in the experiment. We estimate that the supercoupling transmission loss may be practically reduced below 1.9 dB, if the middle channel were manufactured with a steel wall thickness of approximately 7.5 mm or higher (see Fig. 4a).

We can describe the supercoupling phenomenon as a dispersive impedance matching condition, which occurs when the coupling channel (with smaller characteristic impedance than the input waveguide) has an input impedance that appears nearly infinitely stiff. At this matching condition, the phase velocity approaches infinity, as long as $S_2/S_1$ is sufficiently large. Under this condition, we achieve full amplitude transmission and total conservation of the phase, independent of the height and length of the coupling channel. Moreover, our results show that a hard-wall waveguide, when driven near the cut-off frequency of a higher-order mode, exhibits compressibility-near-zero effective material properties, and may be thought of as consisting of two effective soft boundaries, along which the pressure field is equal to zero and the uniform phase of the tunneling mode flips by π. Quite surprisingly, it is possible to suppress the excitation of all other modes in the waveguide, including the dominant plane-wave mode, thanks to the largely mismatched cross-section at the connecting interfaces. We envision a wide range of applicability of this phenomenon, for use in acoustic sensing [35], for the tailoring of acoustic radiation patterns [36]-[37], for acoustic lensing [38], and for enhanced acoustic nonlinearities and sound-matter interactions [14],[33].

**Materials and Methods**

**Transmission-line model for acoustic supercoupling.** Suppose two identical waveguides, each with cross sectional area $S_1$, filled with a fluid with characteristic acoustic impedance $Z_1$. These waveguides are connected as an input and output to an intermediate rectangular acoustic channel, as in Fig. 1, with length $L$, fluid with characteristic impedance $Z_2$ and cross-sectional area $S_2$. Using transmission-line theory, the reflection coefficient for a plane wave from one port of this structure is written as

$$\Gamma = \frac{Z_1 - Z_{in}}{Z_1 + Z_{in}} \qquad (4)$$

where $Z_{in}$ is the impedance seen from the input waveguide when looking into the channel and it is calculated using

$$Z_{in} = Z_2 \frac{Z_1 + jZ_2 \tan(k_z L)}{Z_2 + jZ_1 \tan(k_z L)}. \qquad (5)$$

Plugging Eq. (5) in (4), the reflection coefficient reads as in Eq. (1).

**Cut-off in acoustic waveguides and compressibility-near-zero.** For sound propagating in an acoustic waveguide with hard boundaries filled by a medium with density $\rho_{eff}$ and bulk modulus $\kappa_{eff}$, the following relations can be written

$$\begin{cases} \dfrac{\sqrt{\rho_{eff} \kappa_{eff}}}{S} = Z \\ \sqrt{\dfrac{\rho_{eff}}{\kappa_{eff}}} = \dfrac{k_z}{\omega} \end{cases}, \qquad (6)$$

where $Z$ and $k_z$ are defined as the acoustic impedance and wave vector in the $z$-direction, and $S$ is the cross-sectional area of the waveguide. Solving Eq. (6) for the effective constitutive parameters results in

$$\begin{cases} \rho_{eff} = \dfrac{Z S k_z}{\omega} \\ \kappa_{eff} = \dfrac{Z S \omega}{k_z} \end{cases}, \qquad (7)$$

which allows to retrieve the effective constitutive parameters knowing impedance, wave number, operating frequency and cross-sectional area of the waveguide.

Consider now a waveguide with two parallel soft boundaries at x = {0, a} and two parallel hard boundaries at y = {0, b}. For this configuration, the spatial pressure distribution is given by $p_{mn} = A_{mn} \sin(\frac{m\pi}{a}x)\cos(\frac{n\pi}{b}y)$, and it exhibits cut-off at discrete frequencies

$$f_{mn}^c = \frac{c}{2\pi}\sqrt{(\frac{m\pi}{a})^2 + (\frac{n\pi}{b})^2} . \tag{8}$$

Due to the sinusoidal term in the pressure expression, the mode (m,n) = (0,0) is not supported, resulting in a non-zero cut-off frequency for the dominant mode (m,n) = (1,0), which we denote as the first cut-off frequency, only depending upon the width *a* of the channel.

Momentum conservation requires

$$\nabla p + j\omega\rho_0 u = 0, \tag{9}$$

therefore the particle velocity *u* is given by

$$u = \frac{-A_{mn}e^{-jk_z z}}{j\omega\rho_0}[k_x \cos(k_x x)\cos(k_y y)\hat{x} - k_y \sin(k_x x)\sin(k_y y)\hat{y} - jk_z \sin(k_x x)\cos(k_y y)\hat{z}], \tag{10}$$

and $Z = \frac{p}{S u}$ for the (*m*=1, *n*=0) mode, which gives the result in Eq. (2).

Then, combining Eq. (7) with Eq. (2), we derive the effective material properties of the acoustic waveguide with soft-hard boundaries near the (1,0) mode cutoff, yielding Eq. (3). It is observed that the value of the effective bulk modulus has a pole for $f = \frac{c_0}{2a}$, and consequently the effective compressibility tends to zero.

This CNZ condition can be exploited to induce supercoupling through a soft-hard channel waveguide. These boundaries however can be difficult to realize in practical acoustic media. For a more realistic case, we assume a waveguide configuration in which all boundaries are composed of a hard material. This is a typical scenario for air-filled waveguides. In this case, the spatial pressure distribution is

$$p_{mn} = A_{mn} \cos(\frac{m\pi}{a}x)\cos(\frac{n\pi}{b}y), \tag{11}$$

and the cut-off frequencies are again given by Eq. (8). In this case, the first cut-off frequency is zero, and a plane wave mode can propagate also for very low frequencies in this configuration. However, a compressibility near zero (CNZ) condition can arise when the hard boundary waveguide is operated near a higher cut-off frequency, for instance with (m,n) = (2,0), for which $\kappa_{eff 2} = \frac{\kappa_{0_2}}{1-(\frac{c_0}{af})^2}$, and the CNZ frequency is $f = \frac{c_0}{a}$.

**Numerical modeling.** Finite element analysis was conducted using Comsol Multiphysics. The Pressure Acoustics module was selected with the frequency domain solver. Air was chosen from the Comsol built-in material list as the filling fluid of all structures. Finally, the input and output ports were set to Plane Wave Radiation conditions, while the acoustic source was modeled as an Incident Pressure Field at the input port. In Fig. 1, the walls of the intermediate channel were modeled as either hard or soft boundary conditions. For the simulation of experimental parameters and consideration of vibrational coupling in the results of Fig. 4a, the supercoupoling system was first placed in an external rectangular domain filled with air. This domain enabled the modeling of leakage from the intermediate channel, and was bounded by Perfectly Matched Layers (PMLs) to realize non-reflecting boundaries. Materials were chosen from the built-in material library as Steel AISI 4340 for the intermediate channel walls and aluminum for the input/output channel. The effects of visco-thermal acoustic boundary-layer loss were modeled by specifying the input/output waveguides as coupled to Narrow Region Acoustics. Then, walls in the intermediate channel were numerically modeled as thin elastic shells in the Comsol Acoustic-Shell Interaction Module. Finally, the reflection and transmission coefficient was calculated using a four-microphone measurement technique similar to [40], where the value of complex $k_z$ was derived from numerical simulations using complex sound speed.

**Measurements.** The experimental setup was built from an off-the-shelf steel welder's tool box with dimensions of a = 0.450 m, b = 0.382 m, L = 0.79 m, and wall thickness = 1.54 mm. The input and output waveguides were nearly-identical aluminum tubes with inside diameter = 12.6 mm, length of 92 cm, and thickness of 1.6 mm. The input waveguide was fed by a horn that was mounted transversely to the direction of propagation, and both input and output waveguides were terminated with anechoic foam to suppress standing waves in the tubes. Measurements were carried out with a procedure similar to [40] and following the standards of [41] and [42]. Due to the small dimensions of the input and output waveguides, a modest amount of acoustic boundary-layer loss was observed in the waveguides alone. This was corrected for by employing a complex value of $k_z = \beta + j\alpha$ in the transfer-matrix equations, with $\alpha \approx -0.13 \, \text{m}^{-1}$ according to [43].

**Acknowledgments**

We acknowledge useful discussions with Dr. Michael Haberman, Dr. Dimitrios Sounas, Dr. Mark Hamilton, Dr. Caleb Sieck, and Mr. Anthony Bonomo.

**Funding:** ASEE SMART Scholarship and Swiss National Science Foundation's (SNSF) Doctoral Mobility Fellowship Award under decision Number P1ELP2_165148.

**Author contributions:** HE initiated the research. HE and MB simulated the device and MB and MM performed the experiments. MB and HE conducted the analysis. AA supervised the research. All authors wrote the manuscript.

**Competing interests:** All authors have no competing interests.

**Data and materials availability:** The numerical and experimental data are available upon request.

**Figures**

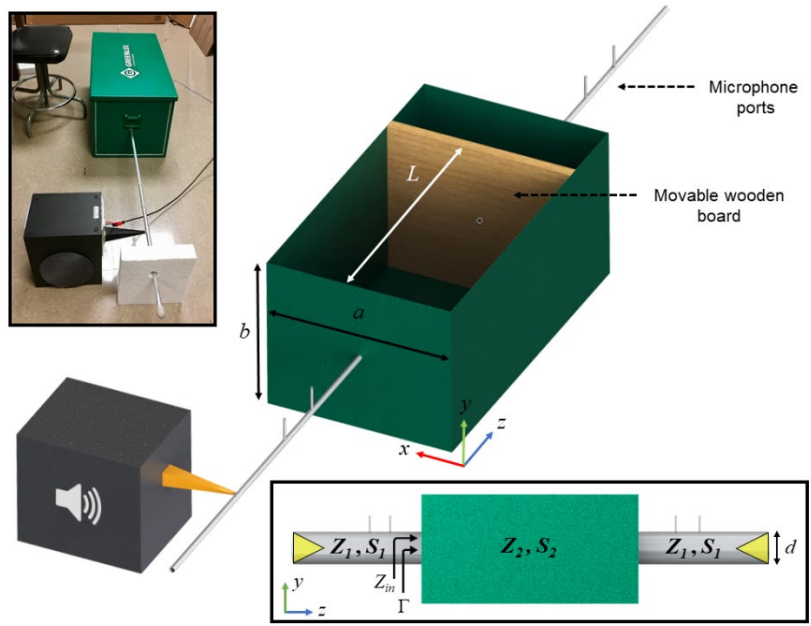

**Fig. 1. Geometry of the CNZ supercoupling experiment.** The geometry under analysis consists of input and output waveguides connected by a much thicker intermediate channel with variable length. (*Inset bottom*): transmission line model. (*Inset top-left*): photograph of the experimental setup.

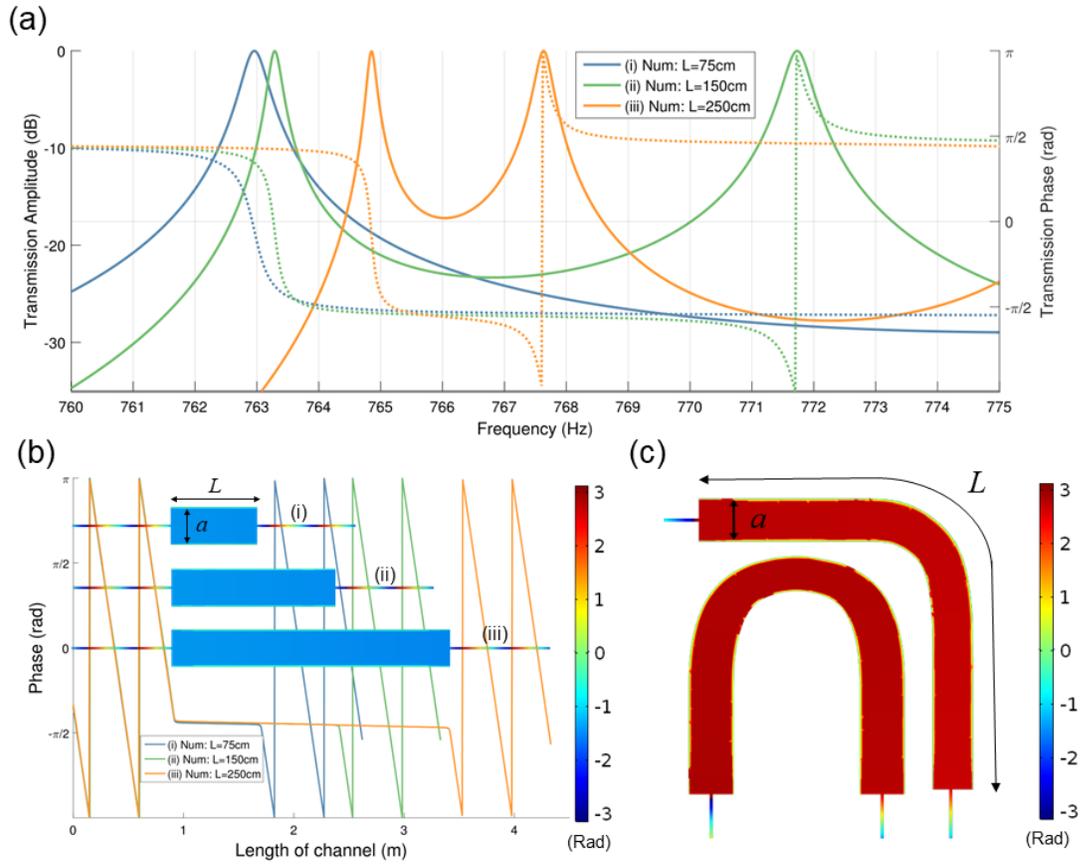

**Fig. 2. Waveguides with two soft and two hard boundaries, operated near the cut-off frequency**. **a.** Calculated transmission phase and amplitude. The design parameters are similar to the experimental setup, with only the length being modified ($a$ = 0.225 m, $b$ = 0.382 m, and $d$ = 12.6 mm). **b.** Phase distribution through the CNZ channel near the first cut off frequency for different lengths. $f_{tunneling}$ (i) = 763.0 Hz, $f_{tunneling}$ (ii) = 763.3 Hz, $f_{tunneling}$ (iii) = 764.9 Hz. **c.** Spatial variation of phase at the CNZ tunneling frequency for waveguides with 90° and 180° bends. We observe uniform phase despite bending of the channel. Here, $f_{tunneling\_90}$ (i) = 756.92 Hz, $f_{tunneling\_180}$ (ii) = 756.76 Hz.

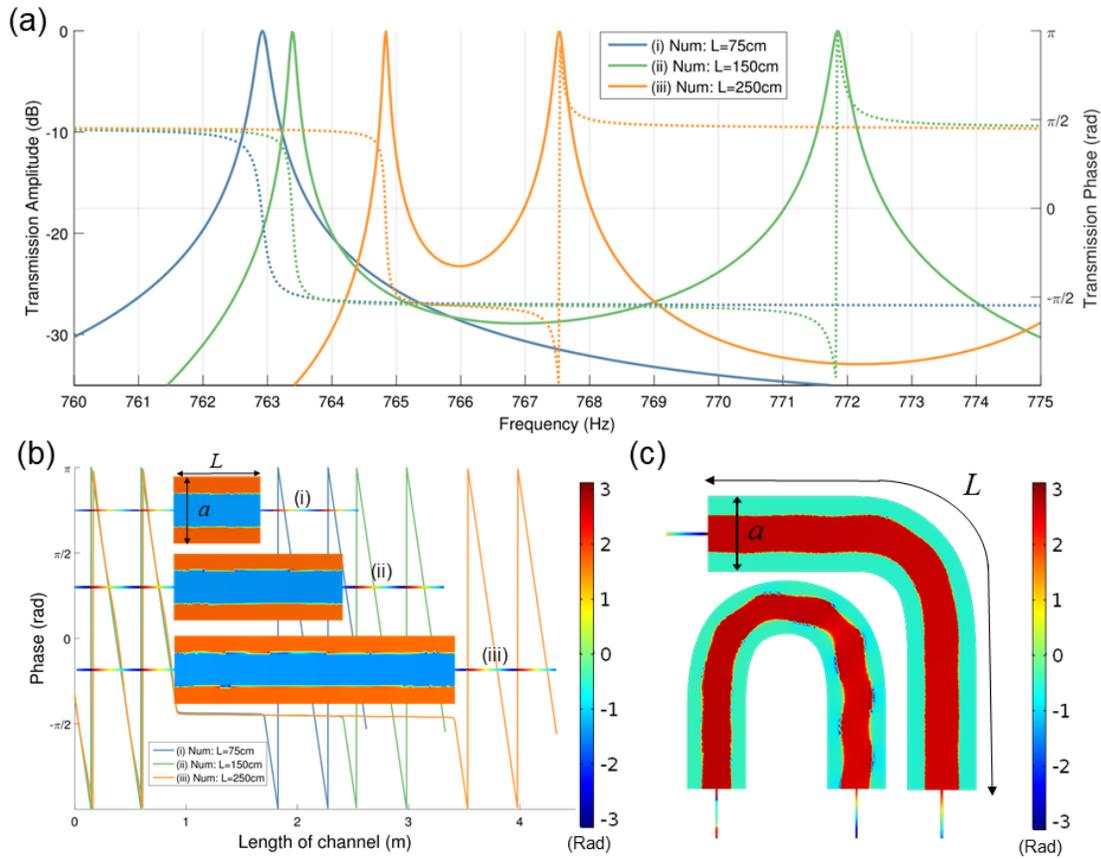

**Fig. 3. Waveguides with hard boundary walls, operated near the higher-order *(2,0)* mode cut-off of the intermediate channel.** We can see that the higher-order mode matches the phase pattern of a soft-hard *(1,0)* mode (from *Fig. 2*) in the center of the channel. This presents an "*effective soft boundary*" along the planes where the phase flips by π. This flip results from a change in sign of the pressure, which is a purely real-valued standing wave. **a.** Numerical results for transmission phase and amplitude. The design parameters are similar to the experimental setup, with only the length being modified ($a$ = 0.450 m, $b$ = 0.382 m, and $d$ = 12.6 mm). **b.** Phase distribution through the CNZ channel near the *(2,0)* cut-off frequency for hard-hard configurations of different lengths. For bottom figure: $f_{tunneling}$ (i) = 762.9 Hz, $f_{tunneling}$ (ii) = 763.4 Hz, $f_{tunneling}$ (iii) = 764.8 Hz. **c.** Spatial variation of phase at the CNZ tunneling frequency under 90° and 180° bends. Here, $a$ = 0.450 m, $b$ = 0.382 m, $f_{tunneling}$ (i) = 764.2 Hz, $f_{tunneling}$ (ii) = 758.97 Hz.

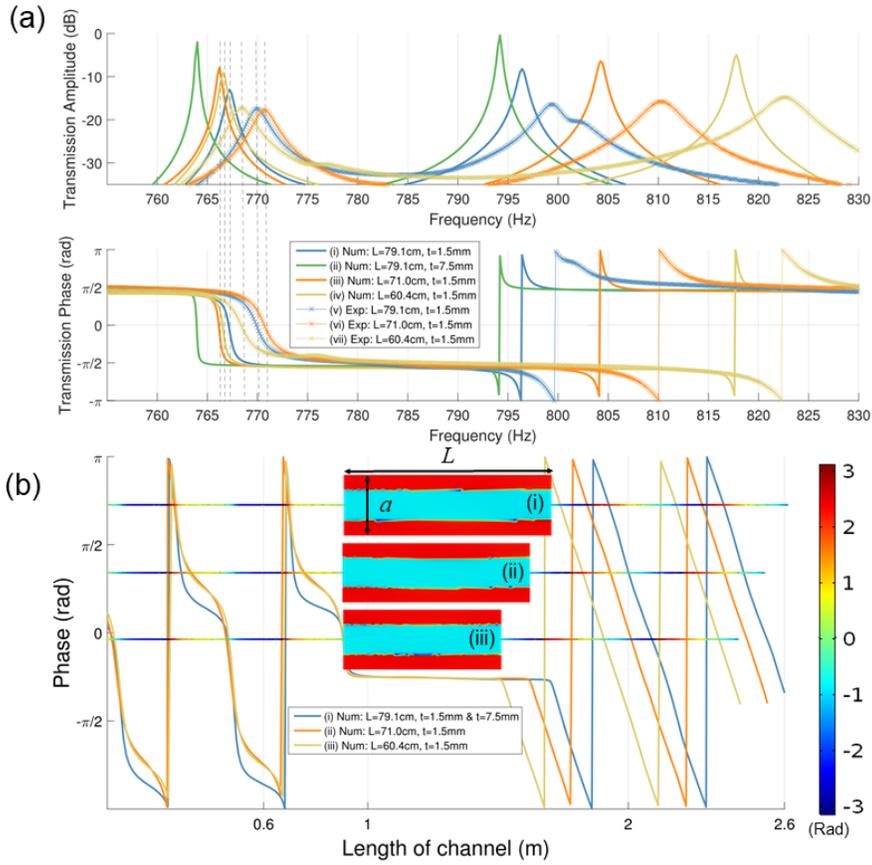

**Fig. 4. Experimental measurements and numerical results for supercoupling channel. a.** Numerical and experimental comparison for the transmission phase and amplitude. Radiation and material loss from the intermediate channel were modeled using elastic shell boundary conditions with a loss tangent of 0.01. Note that curve (ii), in red, shows the transmission amplitude at the tunneling frequency (~765 Hz) is much higher than the others, due to the use of thicker walls for the intermediate channel. The experimental parameters were $a = 0.450$ m, $b = 0.382$ m, and $d = 12.6$ mm. **b.** Spatial phase distribution through the CNZ channel at the tunneling frequency for configurations with different lengths and $f_{tunneling}$ (i) = 767.1 Hz, $f_{tunneling}$ (ii) = 766.2 Hz, $f_{tunneling}$ (iii) = 766.6 Hz.